\documentclass[11pt]{article}
\usepackage[latin9]{inputenc}
\usepackage{geometry}
\usepackage{float}
\usepackage{units}
\usepackage{amsmath}
\usepackage{amsthm}
\usepackage{amssymb}
\usepackage{graphicx}
\usepackage{geometry}
\usepackage{authblk}

\makeatletter
 \theoremstyle{definition}
 \newtheorem*{defn*}{\protect\definitionname}

\@ifundefined{date}{}{\date{}}
\def\be{\begin{equation}}
\def\ee{\end{equation}}
\def\bma{\begin{mathletters}}
\def\ema{\end{mathletters}}

\def\0{\overline{0}}
\def\tr{\mbox{tr}}

\def\q0{\underline{0}}

\def\id{{\mathbb I}}

\def\R{\mathbb{R}}

\def\tr{\mbox{tr}}
\def\one{\leavevmode\hbox{\small1\normalsize\kern-.33em1}}

\def\bra#1{\langle#1|} \def\ket#1{|#1\rangle}

\makeatother

\providecommand{\definitionname}{Definition}

\begin{document}

\title{Convex separation from convex optimization for large-scale
problems}

\author[1]{S. Brierley}
\author[2]{M. Navascu\'es}
\author[3]{T. V\'ertesi }

\affil[1]{\it{DAMTP, Centre for Mathematical Sciences, University of Cambridge, Cambridge CB3 0WA, United Kingdom}}
\affil[2]{\it{Institute for Quantum Optics and Quantum Information (IQOQI), Boltzmangasse 3, 1090 Vienna, Austria}}
\affil[3]{\it{Institute for Nuclear Research, Hungarian Academy of Sciences, H-4001 Debrecen, P.O. Box 51, Hungary}}

\maketitle

\begin{abstract}
We present a scheme, based on Gilbert's algorithm for quadratic minimization [SIAM J. Contrl., vol. 4, pp. 61-80, 1966], to prove separation between a point and an arbitrary convex set $S\subset\R^{n}$ via calls to an oracle able
to perform linear optimizations over $S$. Compared to other methods, our scheme has almost negligible memory requirements and the number of calls to the optimization oracle does not depend on the dimensionality $n$ of the underlying space. We study the speed of convergence of the scheme under different promises on the shape of the set $S$ and/or the location of the point, validating the accuracy of our theoretical bounds with numerical examples. Finally, we present some applications of the scheme in quantum information theory. There we find that our algorithm out-performs existing linear programming methods for certain large scale problems, allowing us to certify nonlocality in bipartite scenarios with upto $42$ measurement settings. We apply the algorithm to upper bound the visibility of two-qubit Werner states, hence improving known lower bounds on Grothendieck's constant $K_G(3)$. Similarly, we compute new upper bounds on the visibility of GHZ states and on the steerability limit of Werner states for a fixed number of measurement settings.
\end{abstract}

\section{Introduction}
\label{intro}

Optimizing a linear function over a convex set $S$ is a fundamental problem that has been studied extensively due to its many practical applications.  A less well investigated, but related question, is the separation problem: given a point, conclude that it is either contained in $S$ or, if not, derive a separating hyperplane. The two problems are polynomially equivalent \cite{grotschel+}, but this reduction does not result in a practical algorithm for the separation problem as it uses the ellipsoid method. Previous fast algorithms for the separation problem have appeared when the convex set has a particular structure, such as in fast cutting plane algorithms for the traveling salesman problem \cite{crowder+80} or perfect matchings  \cite{cunningham84}.

Separation problems frequently arise in quantum information theory. Consider, for example, the problem of deciding whether a vector of experimental data belongs to the set of all hidden variable theories \cite{bell64}. Whilst this set is easily described as the convex hull of all deterministic classical strategies, finding
a separating hyperplane (called a \emph{Bell inequality} \cite{bell64}) is a hard problem \cite{bellNPhard}. Not surprisingly, the best numerical methods available only allow us to tackle small instances of it. The search for new Bell inequalities was our original motivation and we will return to it in Section \ref{sec:Applications}; where we are able to significantly outperform previous approaches based on linear programming methods. 

We now define the separation and optimization problems more carefully.
Given a point $\bar{r}\in R^{n},$ the separation problem asks if
$\bar{r}$ is in $S$ and if not, to find a separating hyperplane.
\begin{defn*}
Weak separation (WSEP): Given a point $\bar{r}\in R^{n}$ and $\delta>0,$
either find $\bar{s}\in S$ such that $\|\bar{r}-\bar{s}\|<\delta$ or if not, find a vector $\bar{c}\in R^{n}$
such that $\bar{c}\cdot\bar{r}>\max\{\bar{c}\cdot\bar{s}|\bar{s}\in S\}.$
\end{defn*}
\noindent Strong separation corresponds to the case $\delta=0.$
We also define the following optimization problem that asks for a
point in $S$ maximixing a linear functional.
\begin{defn*}
Weak optimization (WOPT): Given a vector $\bar{c}\in R^{n}$ and $\delta>0$
find $\bar{r}\in S$ such that $\bar{c}\cdot\bar{r}>\max\{\bar{c}\cdot\bar{x}|\bar{x}\in S\}-\delta.$
\end{defn*}
\noindent Again, \emph{strong} optimization means that $\delta=0.$

The description of the convex set, $S$, has a big impact on the difficulty
of solving either problem. For example, suppose we are presented with
the polytope $S=\{\bar{s}\;|\;\bar{a}_{i}\cdot\bar{s}\leq b_{i}\}$
then the separation problem is easy. Given $\bar{r}\in R^{n}$, test
each of the inequalities $\bar{a}_{i}\cdot\bar{r}\leq b_{i}$; if
they are all satisfied we know that $\bar{r}\in S$ otherwise we can
construct a hyperplane from the inequality that fails. However, to
solve the optimization problem we need to use a linear program. Whilst
this runs in polynomial time, in practice, it can be difficult in
large dimensions. Conversely, if $S$ is presented as the convex hull
of points $S=conv(\{\bar{s}_{1},\ldots,\bar{s}_{n}\})$ optimization
is now easy; we simply evaluate the function at the points $\{\bar{s}_{1},\ldots,\bar{s}_{n}\}$
and take the maximum value. Membership and separation on the other
hand, are harder. Prior to this work, the best approach again involves
running a linear program.

Given that the description of $S$ determines the practical difficulty
in solving either problem, one might be tempted to convert the presentation
of $S$ depending on the problem at hand. Unfortunately the size of
the representation may increase exponentially when converting from
a polytope to a convex hull and vice-versa.

There is, in fact, a method for reducing Separation to optimization
(and vise-versa) such that the number of calls to the optimization
(or separation) oracle is bounded by a polynomial \cite{grotschel+}.
More precisely, denote the dimension of the underlying space as $n$,
the maximum distance between any two elements of $S$ (the diameter) as $D$, and the
desired accuracy $\delta$, then we can solve the weak separation
(optimization) problem with
\[
O\left(\mathrm{poly}\left(n,\log\left(\frac{D}{\delta}\right)\right)\right)
\]
calls to an oracle that solves the strong optimization (separation)
problem.

The usual proof of polynomial equivalence is based on the ellipsoid
method and so does not result in a practically efficient algorithm
\cite{grotschel+}. Ioannou et al. give an improved reduction of separation
from optimization using geometrical reasoning that resulted in the
same Oracle complexity but a more practical algorithm \cite{Ioannou+06}.

In this work, we propose to reduce WSEP to OPT via a third optimization problem:
\begin{defn*}
Weak minimum Distance (WDIST): Given a point $\bar{r}\in R^{n}$ and $\delta>0$,
find $\bar{s}\in S$ such that $\|\bar{r}-\bar{s}\|\leq \mbox{dist}(S,\bar{r})+\delta$.
\end{defn*}
As we will argue, from any point $\bar{s}\in S$ at a nearly-minimum distance with respect to $\bar{r}$, we can either certify that $\|\bar{r}-\bar{s}\|<\delta$ or call OPT one more time to derive a witness $\bar{c}\in\R^n$ certifying that $\bar{r}\not\in S$.

To solve WDIST, we will make use of the algorithm conceived by Gilbert in 1966 \cite{Gilbert} for minimizing quadratic forms in a convex set. In essence, this iterative algorithm generates, via calls to an oracle for strong optimization, a sequence of points $(\bar{s}_k)_k\subset S$ with decreasing distances $d_k$ with respect to $\bar{r}$. When compared to similar reductions, Gilbert's algorithm has two advantages: a) its memory requirements are quite low; b) the number of calls to the oracle does not depend on the dimensionality of $S$. These two features make it an ideal tool to study a variety of scenarios in quantum information science where solving WSEP via convex optimization is too demanding (usually, in terms of computer memory), but nonetheless effective implementations of OPT exist.

In the following, we describe Gilbert's algorithm, and suggest suitable modifications to boost its speed of convergence while keeping the memory requirements low. We then show how the modified Gilbert's algorithm allows a user with a normal desktop to detect non-locality in bipartite scenarios with \emph{up to 42 settings per party}. In connection with this, we will derive new bounds for the nonlocality of Werner states, hence improving known lower bounds for Grothendieck's constant $K_G(3)$. Similarly, we will improve the noise threshold for the steering of isotropic states. Independently of our work, Montina and Wolf also considered the computational cost of of discriminating non-local correlations \cite{montina+16}. The authors also use Gilbert's algorithm and an approximation algorithm to resolve large scale problems as we do. 

Before presenting the applications, we study the speed of convergence of Gilbert's algorithm both theoretically and numerically. We find that, although in worst-case-scenarios the convergence to the optimal distance $d^*$ does scale as $O(\frac{1}{\sqrt{k}})$, as predicted by Gilbert, a promise on the location of the point $\bar{r}$ or on the shape of the set $S$ can boost this scaling, sometimes dramatically. More specifically:

\begin{enumerate}
\item
When $\bar{r}$ is at a distance $d^*>0$ from the set $S$, $d_k-d^*\leq O\left(\frac{D^2}{(d^*)^2k}\right)$.
\item
When $\bar{r}$ is in the interior of $S$, $d_k\leq \mu^k$, for $\mu=1-O\left(\frac{g^2}{D^2}\right)$, where $g$ denotes the distance of $\bar{r}$ to the boundary of $S$.
\item
When $S$ is a curved set (in a sense explained below) and $\bar{r}\not\in S$, then $d_k-d^*\leq \mu^k$, for $\mu=1-O\left(\frac{d^*}{D+1/R}\right)$, where $R$, defined in Section \ref{sec: curved_sets}, denotes the curvature of $S$.
\item
When $S$ is a curved set and we do not have a promise on the value of $d^*$, $d_k-d^*\leq O\left(\frac{M}{k}\right)$, where $M$ depends on the curvature and the diameter of $S$.
\end{enumerate}
Put together, these results imply that WSEP can be solved with negligible memory requirements with a number of oracle calls $O\left(D^4/\delta^4\right)$ for arbitrary sets, and $O\left(D/\delta\right)$ for curved sets.

Inspired by our analysis of convergence, we also present a simple modification to Gilbert's algorithm in which points returned by the optimazation Oracle are stored in memory. We provide numerical evidence that this can produce an exponential boost of convergence when $\bar{r}\not\in S$ - something that is significant in our applications (cf. Table \ref{steeringmemory}).

\section{Description of Gilbert's algorithm}
\label{donJose}

In this section, we describe a scheme for solving the separation
problem for a point $\bar{r}\in R^{n}$ relative to a convex set $S$.
Armed with an oracle for solving (strong) optimization over $S,$
Gilbert's algorithm identifies $\bar{s}^{*}\in S$ with minimum distance
to $\bar{r}$. If $||\bar{r}-\bar{s}^{*}||_{2}<\delta$,
then we have solved WSEP.  In the case $||\bar{r}-\bar{s}^{*}||_{2} \geq \delta$, we have that for any $\bar{s}\in S$, the function $f(\epsilon)=(\bar{r}-\epsilon\bar{s}+(1-\epsilon)\bar{s}^*)^2$ must have positive derivative, $0\leq f'(0)=2(\bar{r}-\bar{s}^*)\cdot (\bar{s}^*-\bar{s})$ since otherwise, there would exist a point closer to $\bar{r}$ than $\bar{s}^*$. Since $(\bar{r}-\bar{s}^*)(\bar{r}-\bar{s}) + (\bar{r}-\bar{s}^*)(\bar{s}-\bar{s}^*) = (\bar{r}-\bar{s}^*)^2$, it follows that
\begin{equation}
(\bar{r}-\bar{s}^{*})\cdot(\bar{r}-\bar{s})\geq(\bar{r}-\bar{s}^{*})^{2}\geq \delta.\label{eq: witness}
\end{equation}
Thus the vector $\bar{c}=\bar{r}-\bar{s}^{*}$ witnesses $\bar{r}\notin S$ (note that, if $\bar{r}\in S$, $\min_{\bar{s}}\bar{c}\cdot (\bar{r}-\bar{s})\leq 0$).
In fact, $\bar{c}$ is, in a sense, the optimal witness, and applying the stopping
criteria given in Appendix~B 
%\ref{sec:App: Stopping} 
terminates the algorithm sooner.

\paragraph{Gilbert's algorithm} The algorithm presented by Gilbert in reference \cite{Gilbert} to solve WDIST is as follows:
\begin{enumerate}
\item Choose any $\bar{s}_{0}\in S$.
\item Given $\bar{s}_{k}\in S$, use the oracle to maximize the overlap $(\bar{r}-\bar{s}_k)\cdot \bar{s}$ over all $\bar{s}\in S$. That is, find the point $\bar{s}'_{k}$
such that
\begin{equation}
(\bar{r}-\bar{s}_{k})\cdot(\bar{r}-\bar{s}'_{k})\leq(\bar{r}-\bar{s}_{k})\cdot(\bar{r}-\bar{s}),\label{eq:Step 2}
\end{equation}
for all $\bar{s}\in S$.
\item Calculate the convex combination of $\bar{s}_{k}$, $\bar{s}_{k}'$
that minimizes the distance with respect to $\bar{r}$. In other words,
solve the optimization problem:

\begin{equation}
\min_{0\leq\epsilon\leq1}\|(1-\epsilon)\bar{s}_{k}+\epsilon\bar{s}_{k}'-\bar{r}\|_{2}.\label{eq: Step 3 prime}
\end{equation}

\noindent The value that minimizes Eq. (\ref{eq: Step 3 prime})
\begin{equation}
\epsilon_{k}\equiv\min\left\{\frac{(\bar{r}-\bar{s}_{k})\cdot(\bar{s}'_{k}-\bar{s}_{k})}{(\bar{s}'_{k}-\bar{s}_{k})\cdot(\bar{s}'_{k}-\bar{s}_{k})},1\right\}
\label{formulita_epsi}
\end{equation}
 identifies the next point in the algorithm, $\bar{s}_{k+1}\equiv(1-\epsilon_{k})\bar{s}_{k}+\epsilon_{k}\bar{s}_{k}'$.

\item Go to step 2.
\end{enumerate}

\begin{figure}[h]
\centering \includegraphics[width=9cm]{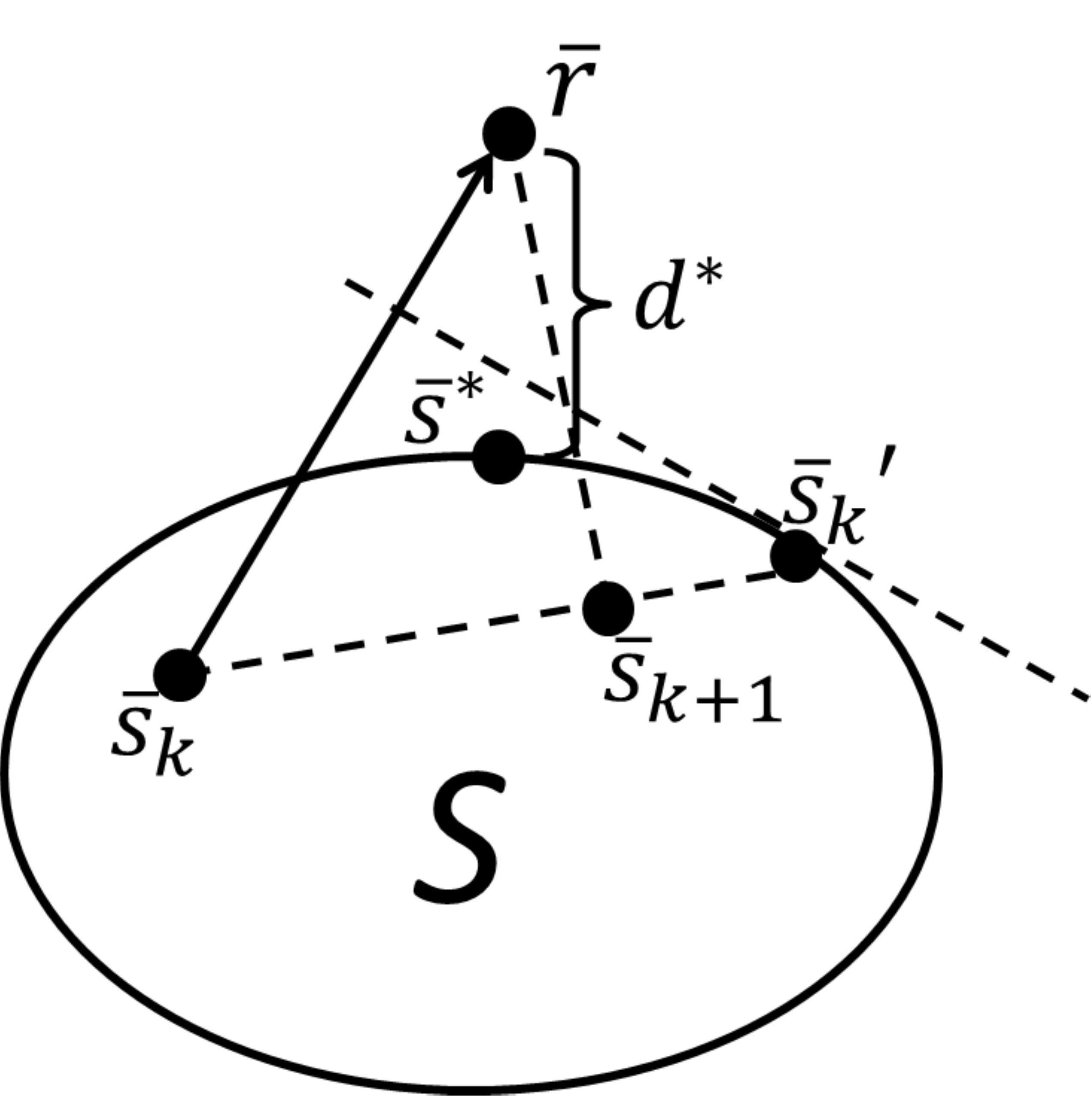} \caption{Geometrical description of Gilbert's algorithm.}
\label{don_jose}
\end{figure}

\noindent A depiction of a single iteration of Gilbert's algorithm in 2 dimensions is given in Fig \ref{don_jose}

\paragraph{Gilbert's algorithm with memory}
While it can be shown (below) that Gilbert's algorithm always converges to the optimal distance, sometimes it does so with the lowest possible speed $O(1/ \sqrt{k})$. We have observed that, in some of these instances, the speed of convergence is boosted if we keep a list $L_{opt}$ of the last $m$ points returned by the strong optimization oracle, and find the convex combinations of them and $\bar{s}_{k}$ that minimizes the distance with respect to $\bar{r}$. An intuition for why this is a good idea can be seen by considering how the algorithm converges when $S$ is a rectangle (see Fig \ref{Fig: rectangle}) where convergence for Gilbert's algorithm is $O(1/k)$ but by storing a list $L_{opt}$, of size 2, convergence is achieved in $O(1)$ iterations.

We call the following modification, \emph{Gilbert's algorithm with memory}:
\begin{enumerate}
\item Choose any $\bar{s}_{0}\in S$.
\item Given $\bar{s}_{k}\in S$, use the oracle to maximize the overlap $(\bar{r}-\bar{s}_k)\cdot \bar{s}$ over all $\bar{s}\in S$. Add $\bar{s}'_{k}$ to $L_{opt}$. If $L_{opt}$ contains $m+1$ entries, erase the first one.
\item Project $\bar{r}$ onto the set $L_{opt}\cup\{\bar{s}_{k}\}$. In
other words, solve the linear least squares with constraint problem:
\begin{align}
\min_{\bar{x}}\qquad & \|A\bar{x}-\bar{r}\|_{2}\nonumber \\
\mathrm{\textrm{subject to}\qquad} & \bar{x}\succeq 0\label{eq: Step 3}\\
 & (1,\cdots, 1)\cdot \bar{x}=1\nonumber
\end{align}
where the elements of $L_{opt}\cup\{\bar{s}_{k}\}$ are arranged as
columns in $A$. Set the next point in the algorithm as the solution
to Eq. (\ref{eq: Step 3}), $\bar{s}_{k+1}\equiv A\bar{x}_{min}$.
\item Go to step 2.
\end{enumerate}
In practice, storing a large number of optimization points makes step
3 time consuming. In fact, the case $L_{opt} = S = conv(\{\bar{s}_{1},\ldots,\bar{s}_{n}\})$ corresponds to the original weak minimum distance problem. In the implementations discussed in Sections \ref{sec: Numerics}
and \ref{sec:Applications}, we balance $m$ to trade-off the number of
iterations and the time taken for each step.

\paragraph{Gilbert's algorithm with a heuristic oracle}
For some convex sets $S$, the oracle solving OPT is too time consuming. In such cases, a cheaper heuristic algorithm is very advantageous. If the approximate OPT oracle guarantees to return a point, $\bar{s}'_{k}\in S$ we have the obvious modification of Gilbert's algorithm, where Step 2 is replaced by

\begin{enumerate}

\item[2'] Given $\bar{s}_{k}\in S$, use the heuristic to maximize the overlap $(\bar{r}-\bar{s}_k)\cdot \bar{s}$ over all $\bar{s}\in S$.

\end{enumerate}

If the (decreasing) sequence of distances $(d_k)_k$ gets very close to zero, then we have solved WSEP. If, on the contrary, $(d_k)_k$ seems to converge to $d^*>0$, we can invest time resources and invoke the exact oracle \emph{once} to maximize the overlap with $\bar{c}=\bar{r}-\bar{s}_k$. Because of the considerations exposed above, if $s_k$ is sufficiently close to the optimal value, then $\min_{\bar{s}\in S}\bar{c}\cdot(\bar{r}-\bar{s})>0$, in which case $\bar{r}\not\in S$, as certified by the witness $\bar{c}$.

We use a heuristic oracle when solving the non-locality of two-qubit Werner states problem in Sec \ref{nonlocwerner}.

\section{Analysis of convergence}

In this section we analyze the convergence of Gilbert's algorithm. We consider the case where
we store only a single optimization point, $m=1$ and derive bounds on the convergence for arbitrary sets $S$.

\subsection{Polynomial convergence for arbitrary sets \label{sub:Don-Jose-without}}

Define the vectors $\bar{d}_{k}\equiv\bar{r}-\bar{s}_{k}$, $\bar{d}'_{k}\equiv\bar{r}-\bar{s}'_{k}$
and $\bar{d}^{*}\equiv\bar{r}-\bar{s}^{*}$ and the diameter of the
smallest ball containing $S$ as
\[
D\equiv\sup\{||\bar{s}-\bar{t}||_{2}\;|\;\bar{s},\bar{t}\in S\}.
\]
With this notation, the linear inequality at step 2 can be rewritten
as $\bar{d}_{k}\cdot\bar{d}_{k}'\leq\bar{d}_{k}\cdot\bar{d}$, for
any $\bar{d}=\bar{r}-\bar{s}$, with $\bar{s}\in S$.

Note that

\begin{equation}
\bar{d}_{k}\cdot(\bar{d}_{k}-\bar{d}_{k}')\geq\bar{d}_{k}\cdot(\bar{d}_{k}-\bar{d}^{*})=d_{k}^{2}-\bar{d}_{k}\cdot\bar{d}^{*}\geq d_{k}(d_{k}-d^{*}).\label{funda}
\end{equation}

\noindent Then,

\begin{equation}
\bar{d}_{k+1}=(1-\epsilon_{k})\bar{d}_{k}+\epsilon_{k}\bar{d}_{k}'=\bar{d}_{k}+\epsilon_{k}\bar{v}_{k},\label{defin}
\end{equation}

\noindent for some $0\leq\epsilon_{k}\leq1$, with $\bar{v}_{k}=\bar{d}_{k}'-\bar{d}_{k}$. In Appendix~A we show that the value of $\epsilon_k$ minimizing the norm of the right hand side of the above equation is given by (\ref{formulita_epsi}).

There are two possibilities:
\begin{enumerate}
\item $-\frac{\bar{d}_k\cdot \bar{v}_k}{\bar{v}_k^2}>1$, or, equivalently, $\bar{d}_{k}'\cdot\bar{v}_{k}=\bar{d}_{k}'\cdot(\bar{d}_{k}'-\bar{d}_{k})\leq0$, in which case $\epsilon_{k}=1$. Then,

\begin{equation}
\bar{d}_{k}^{2}-\bar{d}_{k+1}^{2}=\bar{d}_{k}^{2}-(\bar{d}'_{k})^{2}\geq\bar{d}_{k}\cdot(\bar{d}_{k}-\bar{d}_{k}')\geq d_{k}(d_{k}-d^{*}),
\label{epsi1}
\end{equation}

\noindent where the last inequality follows from Eq. (\ref{funda}),
and the previous one is a consequence of the assumption $\bar{d}_{k}'\cdot(\bar{d}_{k}'-\bar{d}_{k})\leq0$.
We can carry further the approximation and conclude that:

\begin{equation}
d_{k}-d_{k+1}\geq\frac{d_{k}(d_{k}-d^{*})}{d_{k}+d_{k+1}}\geq\frac{d_{k}-d^{*}}{2}\geq\frac{(d_{k}-d^{*})^{3}}{2D^{2}}.\label{approx}
\end{equation}

\item $\bar{d}_{k}'\cdot\bar{v}_{k}>0$, in which case

\begin{equation}
\epsilon_{k}=\frac{\bar{d}_{k}\cdot(\bar{d}_{k}-\bar{d}_{k}')}{(\bar{d}_{k}-\bar{d}_{k}')^{2}}.
\end{equation}

\noindent Substituting this value in eq. (\ref{defin}), it can be verified that $\bar{d}_{k+1}\cdot\bar{v}_{k}=0$, and so

\begin{equation}
\bar{d}_{k+1}^{2}=\bar{d}_{k+1}\cdot\bar{d}_{k}+\epsilon_{k}\bar{d}_{k+1}\cdot\bar{v}_{k}=\bar{d}_{k+1}\cdot\bar{d}_{k}.
\end{equation}

\noindent It follows that

\begin{equation}
d_{k}^{2}-d_{k+1}^{2}=\bar{d}_{k}\cdot(\bar{d}_{k}-\bar{d}_{k+1})=\epsilon_{k}\bar{d}_{k}\cdot(\bar{d}_{k}-\bar{d}_{k}')=\frac{[\bar{d}_{k}\cdot(\bar{d}_{k}-\bar{d}_{k}')]^{2}}{(\bar{d}_{k}-\bar{d}_{k}')^{2}}\label{identity}
\end{equation}

\noindent Invoking relation (\ref{funda}), we arrive at

\begin{equation}
d_{k}^{2}-d_{k+1}^{2}\geq\frac{d_{k}^{2}(d_{k}-d^{*})^{2}}{D^{2}}.
\end{equation}

\noindent Therefore,

\begin{equation}
d_{k}-d_{k+1}\geq\frac{d_{k}^{2}(d_{k}-d^{*})^{2}}{D^{2}(d_{k}+d_{k+1})}\geq\frac{d_{k}(d_{k}-d^{*})^{2}}{2D^{2}}\geq\frac{(d_{k}-d^{*})^{3}}{2D^{2}}.\label{approx2}
\end{equation}

\end{enumerate}
\noindent To conclude: no matter what the value of $\bar{d}_{k}'\cdot \bar{v}_k$
is, we have that

\begin{equation}
d_{k}-d_{k+1}\geq\frac{(d_{k}-d^{*})^{3}}{2D^{2}}.
\end{equation}

Now, define $z_{k}\equiv(d_{k}-d^{*})/\sqrt{2}D$. Then, the last
relation reads:

\begin{equation}
z_{k}-z_{k+1}\geq z_{k}^{3}.\label{recu}
\end{equation}

\noindent Since $0<z_{k+1}\leq z_{k}$, we have that

\begin{eqnarray}\label{truqui}
\sum_{j=0}^{k-1}\frac{1}{2}\left(\frac{1}{z_{j+1}^{2}}-\frac{1}{z_{j}^{2}}\right)  &=& \sum_{j=0}^{k-1}\frac{1}{2}\left(\frac{(z_{j}+z_{j+1})(z_{j}-z_{j+1})}{z_{j}^{2}z_{j+1}^{2}}\right) \nonumber \\
& \geq & \sum_{j=0}^{k-1}\frac{z_{j}-z_{j+1}}{z_{j+1}z_{j}^{2}} \nonumber \\
& \geq &\sum_{j=0}^{k-1}\frac{z_{j}-z_{j+1}}{z_{j}^{3}} \\ \nonumber
& \geq & \sum_{j=0}^{k-1}1 = k \, ,
\end{eqnarray}
where we used (\ref{recu}) to derive the last inequality.
The left hand side of relation (\ref{truqui}) is a telescopic series;
its value is equal to $\frac{1}{2}(\frac{1}{z_{k}^{2}}-\frac{1}{z_{0}^{2}})$.
It follows that $z_{k}^{2}\leq\frac{1}{2k+1/z_{0}^{2}}$, that is:

\begin{equation}
d_{k}\leq d^{*}+\frac{\sqrt{2}D}{\sqrt{2k+\frac{2D^{2}}{(d_{0}-d^{*})^{2}}}}\leq d^{*}+\frac{D}{\sqrt{k+1}}.\label{bound_malo}
\end{equation}

\noindent The algorithm thus converges at least as
\begin{equation}
k=O(D^{2}/\delta^{2}).\label{eq: Inside complexity}
\end{equation}

In the case $d^{*}>0$, that is $\bar{r}\not\in S$, we can improve
the latter bound. Indeed, notice that, for $d^{*}>0$ we have

\begin{equation}
d_{k}-d_{k+1}\geq\frac{d_{k}(d_{k}-d^{*})^{2}}{2D(D+d^{*})}\geq\frac{d^{*}(d_{k}-d^{*})^{2}}{2D(D+d^{*})},
\end{equation}

\noindent this follows from Eqs. (\ref{approx}), (\ref{approx2}).
Define $y_{k}\equiv(d_{k}-d^{*})/\Delta$, with $\Delta\equiv2D(D+d^{*})/d^{*}$;
we thus have that $y_{k}-y_{k+1}\geq y_{k}^{2}$. Using similar arguments,
one concludes that $y_{k}\leq\frac{1}{k+1/y_{0}}$, and, consequently,

\begin{equation}
d_{k}\leq d^{*}+\frac{\Delta}{k+\frac{d_{0}-d^{*}}{\Delta}}.\label{bound}
\end{equation}

\noindent If $\bar{r}\not\in S$, the scaling of the algorithm to solve WDIST is
bounded by
\[
k=O\left(\frac{D^{2}}{d^*\delta}\right)
\]

It rests to ascertain how good this method is to solve WSEP. Clearly, if $d^*\leq \delta/2$, by eq. (\ref{eq: Inside complexity}) we have that $k=O(D^2/\delta^2)$ calls to the oracle would be enough to find a point $s$ with $\|\bar{r}-\bar{s}\|_2<\delta$. If, on the contrary, $d^*> \delta/2$, we would like to find a witness to detect that $\bar{r}\not\in S$. In Appendix~B, 
%\ref{sec:App: Stopping},
it is shown that this is possible provided that $d_k-d^*<O\left(\frac{(d^*)^3}{D^2}\right)$. Invoking bound (\ref{bound}), we thus have that $k=O\left(\frac{D^4}{\delta^{4}}\right)$ are required. This last bound hence captures the worst-case general performance of the scheme to solve WSEP($\delta$).

\vspace{10pt}

Let us suppose now that $\bar{r}$ actually belongs to the \emph{interior} of the set $S$, and let $g>0$ be the distance between $\bar{r}$ and the boundary $\partial S$ of $S$. That is:

\be
g\equiv \min_{\bar{s}\in \partial S} \|\bar{r}-\bar{s}\|_2.
\ee

Given $\bar{s}_k$, define $\bar{s}_k(\lambda)\equiv\bar{r}+\lambda \bar{d}_k$, and $\lambda_k\equiv \max\{\lambda:\bar{s}_k(\lambda)\in S\}$. From the equation above, we have that $\lambda_k\geq \frac{g}{d_k}$.

Now, $\bar{d}_k\cdot\bar{d}'_k\leq \bar{d}_k\cdot(\bar{r}-\bar{s}_k(\lambda_k))=-\lambda_k\bar{d}_k^2<0$. It follows that $\bar{d}_k\cdot(\bar{d}_k-\bar{d}'_k)\geq d_k(d_k+g)\geq g d_k$, and so, by (\ref{identity}) we have that, for $\epsilon_k<1$,

\be
d_k^2-d_{k+1}^2\geq \frac{g^2d_k^2}{D^2}.
\ee

\noindent Setting $d^*=0$ in eq. (\ref{epsi1}), we have that the case $\epsilon_k=1$ does not occur, unless $d_{k+1}=0$. In either case, the above expression always holds, and so

\be
d^2_{k+1}\leq \left(1-\frac{g^2}{D^2}\right)d_{k}^2.
\ee

\noindent Iterating, we find that

\be
d_{k}\leq \left(1-\frac{g^2}{D^2}\right)^{k/2}d_0.
\ee

\noindent The convergence of the algorithm for inner points has an exponential boost.

\subsection{Curved sets\label{sec: curved_sets}}

\begin{figure}
\centering \includegraphics[width=9cm]{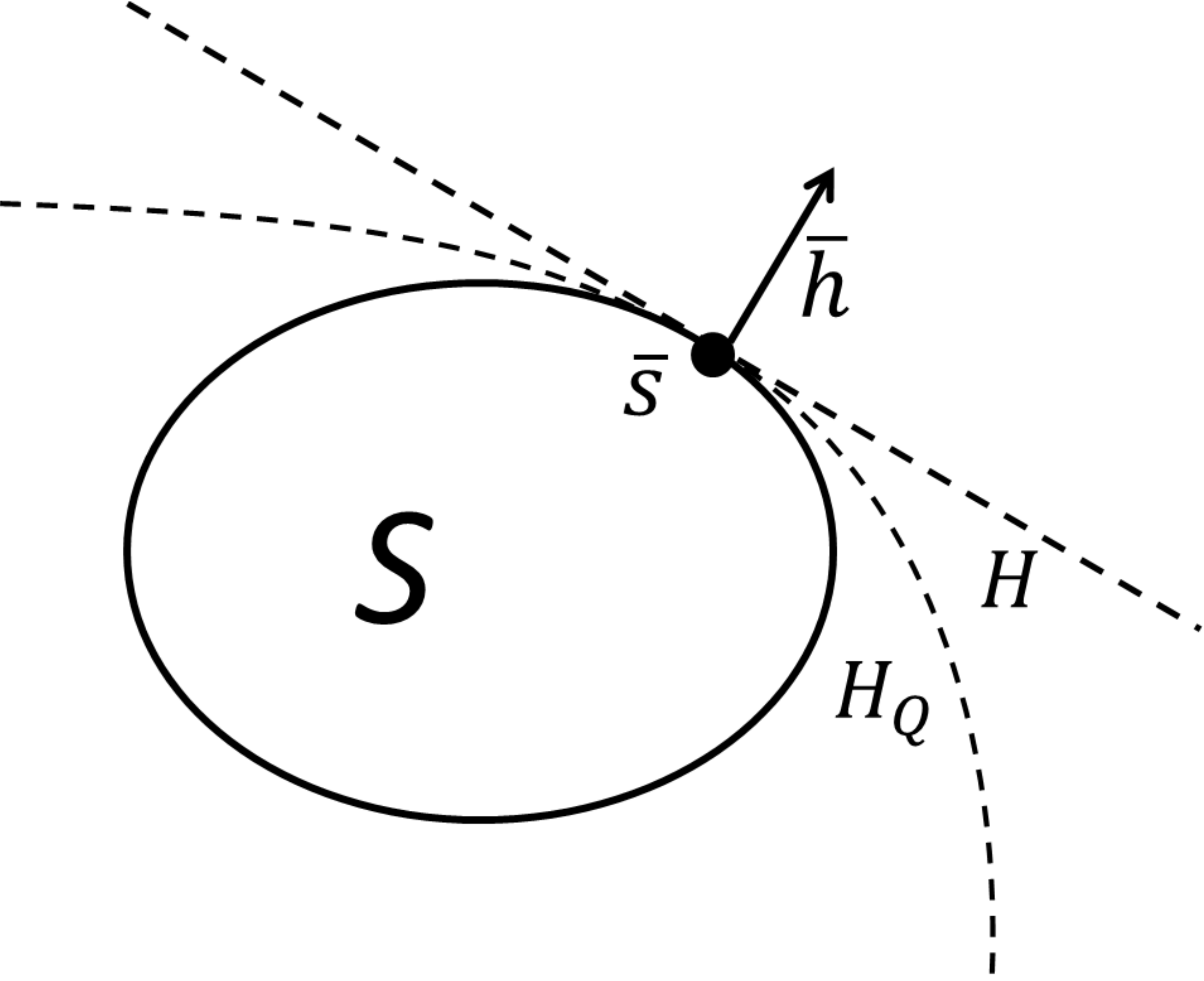} \caption{\textbf{Geometrical meaning of curvature.} The plot depicts a convex set $S$ with a distinguished boundary point $\bar{s}\in S$ with normal vector $\bar{h}$. $\bar{h}$ defines a linear approximation to the set via the inequality $\bar{h}\cdot(\bar{s}'-\bar{s})\leq0$ for all $\bar{s}'\in S$. If the set has a certified curvature $R$, this approximation can be improved via the quadratic witness defined by Eq. (\ref{quadratic}).}
\label{curved_set}
\end{figure}

We now consider the convergence of Gilbert's algorithm when the boundary of $S$
is curved. Let $\bar{s}\in S$ be an arbitrary point on the boundary
of $S$. Then we can always find a \emph{tangent} half-space of the
form $H\equiv\{\bar{x}:\bar{h}\cdot(\bar{x}-\bar{s})\leq0\}$, with
$\|\bar{h}\|=1$, such that $S\subset H$. $H$ can thus be regarded
as a rough outer linear approximation to $S$, see Figure \ref{curved_set}.

We say that $S$ is \emph{curved} if this approximation can be refined
into a quadratic constraint. More specifically, $S$ has curvature
$R$ iff, for any point $\bar{s}$ on the boundary of $S$, with tangent
half-space $H\equiv\{\bar{x}:\bar{h}\cdot(\bar{s}-\bar{x})\leq0\}$,
the set of points

\begin{equation}
H_{Q}\equiv\{\bar{x}:\bar{h}\cdot(\bar{x}-\bar{s})+R(\bar{x}-\bar{s})^{T}\cdot(\id-\bar{h}\bar{h}^{T})\cdot(\bar{x}-\bar{s})\leq0\}\label{quadratic}
\end{equation}
also contains $S$. A set is called \emph{strictly} convex if and only if $R>0$.

Now, suppose that $S$ has curvature $R$, and consider the point
$\bar{s}'_{k}$ derived from $\bar{s}_{k}$ in the second step of
Gilbert's algorithm. Clearly, $\bar{s}'_{k}$ belongs to the boundary. Moreover,
by construction, it has a tangent half-space $H\equiv\{\bar{x}:\frac{\bar{d_{k}}}{d_{k}}\cdot(\bar{x}-\bar{s}'_{k})\leq0\}$.
From our definition of curvature it hence follows that $\bar{s}_{k}\in S$
must belong to the set (\ref{quadratic}), with $\hat{h}=\frac{\bar{d_{k}}}{d_{k}}$.
Now, calling $t\equiv\bar{d}_{k}\cdot(\bar{d}_{k}-\bar{d}_{k}')=\bar{d}_{k}\cdot(\bar{s}_{k}'-\bar{s}_{k})$,
we have that

\begin{equation}
(\bar{d}_{k}-\bar{d}_{k}')^{2}=\frac{t^{2}}{d_{k}^{2}}+(\bar{d}_{k}-\bar{d}_{k}')^{T}\cdot\left(\id-\frac{\bar{d}_{k}\bar{d}_{k}^{T}}{d_{k}^{2}}\right)\cdot(\bar{d}_{k}-\bar{d}_{k}')\leq\frac{t^{2}}{d_{k}^{2}}+\frac{t}{Rd_{k}},
\end{equation}

\noindent where the last inequality follows from Eq. (\ref{quadratic}).

\noindent For the case $\epsilon_k\not=1$, this last relation, together with Eq. (\ref{identity}), implies that

\begin{equation}
d_{k}^{2}-d_{k+1}^{2}=\frac{t^{2}}{(\bar{d}_{k}-\bar{d}_{k}')^{2}}\geq\frac{t^{2}}{\frac{t^{2}}{d_{k}^{2}}+\frac{t}{Rd_{k}}}=\frac{d_{k}^{2}t}{t+\frac{d_{k}}{R}}\geq\frac{d_{k}^{2}t}{d_{k}D+\frac{d_{k}}{R}}\geq\frac{d_{k}^{2}(d_{k}-d^{*})}{D+\frac{1}{R}},
\end{equation}

\noindent the last inequality being a consequence of (\ref{funda}).
We thus have that

\begin{equation}
d_{k}-d_{k+1}\geq\frac{d_{k}^{2}(d_{k}-d^{*})}{(d_{k}+d_{k+1})(D+\frac{1}{R})}\geq\frac{d_k(d_{k}-d^{*})}{2\left(D+\frac{1}{R}\right)}\geq\frac{d^{*}(d_{k}-d^{*})}{2\left(D+\frac{1}{R}\right)}.
\label{curv_rel}
\end{equation}

\noindent For the case $\epsilon_{k}=1$, we have by (\ref{approx}) that

\begin{equation}
d_{k}-d_{k+1}\geq\frac{d_{k}-d^{*}}{2}.
\end{equation}

Defining $z_{k}\equiv d_{k}-d^{*}$, we hence obtain the recursive
relation

\begin{equation}
z_{k}-z_{k+1}\geq\lambda z_{k},
\end{equation}

\noindent with

\begin{equation}
\lambda=\min\left(\frac{1}{2},\frac{d^{*}}{2\left(D+\frac{1}{R}\right)}\right).
\end{equation}

\noindent It is easy to verify that $z_{k}\leq(1-\lambda)^{k}z_{0}$,
and therefore

\begin{equation}
d_{k}\leq d^{*}+(1-\lambda)^{k}(d_{0}-d^{*})\leq d^{*}+(1-\lambda)^{k}D.\label{exponential}
\end{equation}

\noindent If $d^*$ is comparable to $D$ and the curvature $R$ is not so small, we should hence expect Gilbert's algorithm to approach the optimal solution exponentially fast.

For small $d^*/D$, however, the number of iterations to solve WDIST is bounded by
\begin{align}
k & \leq\frac{\log\left(\frac{\delta}{D}\right)}{\log\left(1-\lambda\right)}\\
 & =O\left(\left(\frac{D+\frac{1}{R}}{d^*}\right)\log\left(\frac{D}{\delta}\right)\right).
 \label{num_it_curv}
\end{align}

In the absence of a promise on the value of $d^*$, we can estimate the speed of convergence of Gilbert's algorithm by taking the second inequality of eq. (\ref{curv_rel}) and then approximating $d_k$ by $d_k-d^*$ in the right hand side. We thus find that

\be
d_k-d_{k+1} \geq\frac{(d_{k}-d^{*})^2}{2\left(D+\frac{1}{R}\right)}.
\ee

Similarly as in the previous section, we find that

\be
d_k\leq d^*+\frac{2(D+1/R)}{k+\frac{d_0-d^*}{2(D+1/R)}}.
\ee

\noindent In other words, the number of iterations to solve WDIST is $O\left(\frac{(D+1/R)}{\delta}\right)$.

What about WSEP? If $D\gg d^*>\delta/2$, we find, by Appendix~B 
%\ref{sec:App: Stopping} 
and (\ref{num_it_curv}), that a witness will be produced after $O\left(\frac{D+1/R}{\delta}\log\left(\frac{D}{\delta}\right)\right)$ iterations. If, on the other hand, $d^*<\delta/2$, by the equation above we find that a point at a distance $\delta$ will be found after $O\left(\frac{(D+1/R)}{\delta}\right)$ calls to the oracle. Hence, in worst-case scenarios, the total number of calls is $O\left(\frac{D+1/R}{\delta}\log\left(\frac{D}{\delta}\right)\right)$.

\section{How accurate are the bounds? \label{sec: Numerics}}

\begin{figure}[H]
\centering \includegraphics[width=10cm]{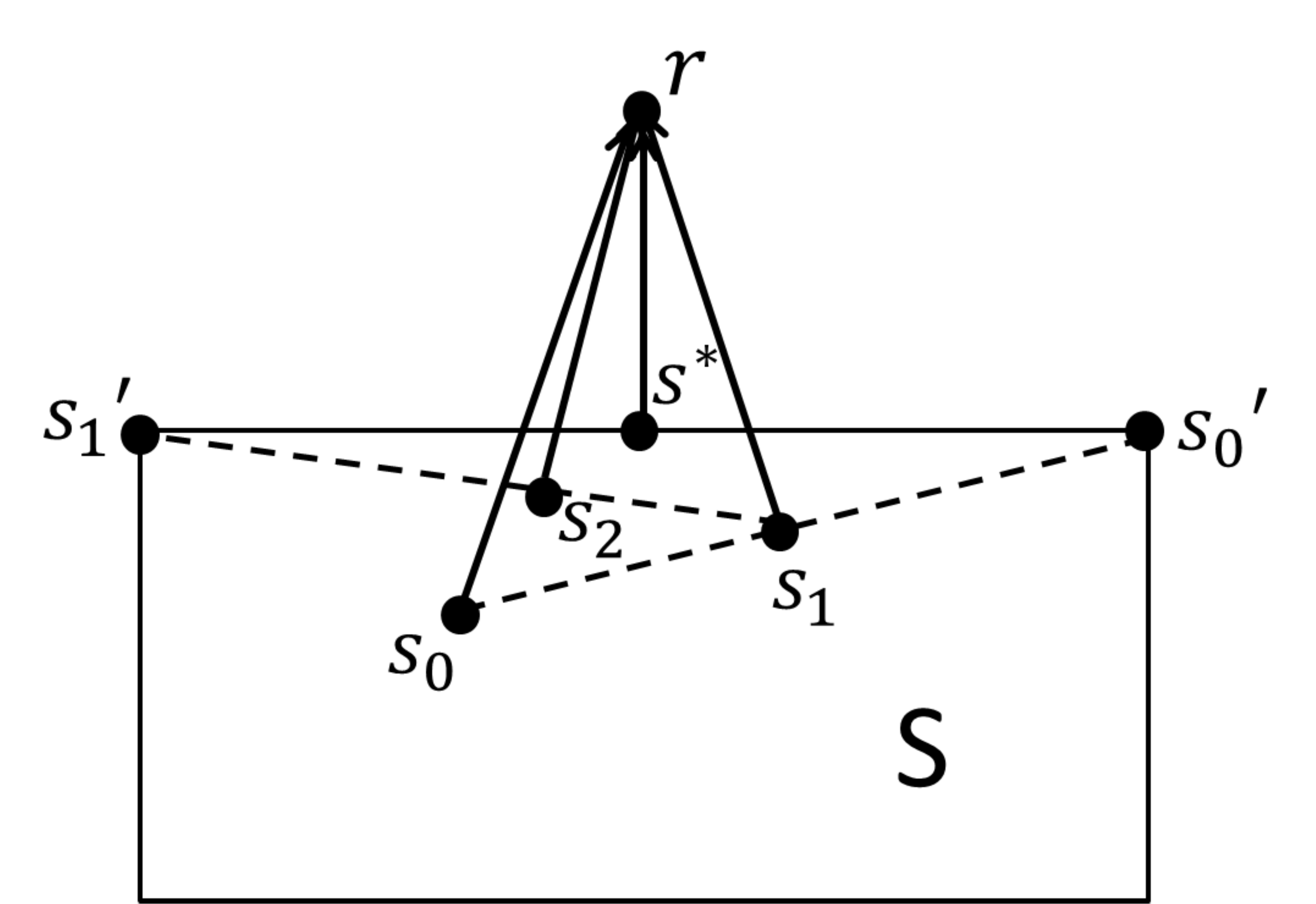} \caption{\textbf{Lower bounds on the speed of convergence.} In the picture, the convex set $S$ is a rectangle. It can be seen that, for all $k$, $s_{2k}'$, $s_{2k+1}'$ are the right and left upper vertices of the rectangle, respectively.}
\label{Fig: rectangle}
\end{figure}

Let us investigate whether the bounds (\ref{bound_malo}), (\ref{bound})
are asymptotically tight. A possible convex set where the speed of
convergence may be computed exactly is the rectangle depicted in Figure
\ref{Fig: rectangle}. Note that in this setting the value of $\epsilon_{k}$
decreases as Gilbert's algorithm approaches $s^{*}$: in view of Eq. (\ref{identity})
it is hence likely that the speed of convergence is low.

\begin{figure}[H]
\centering \includegraphics[width=12cm]{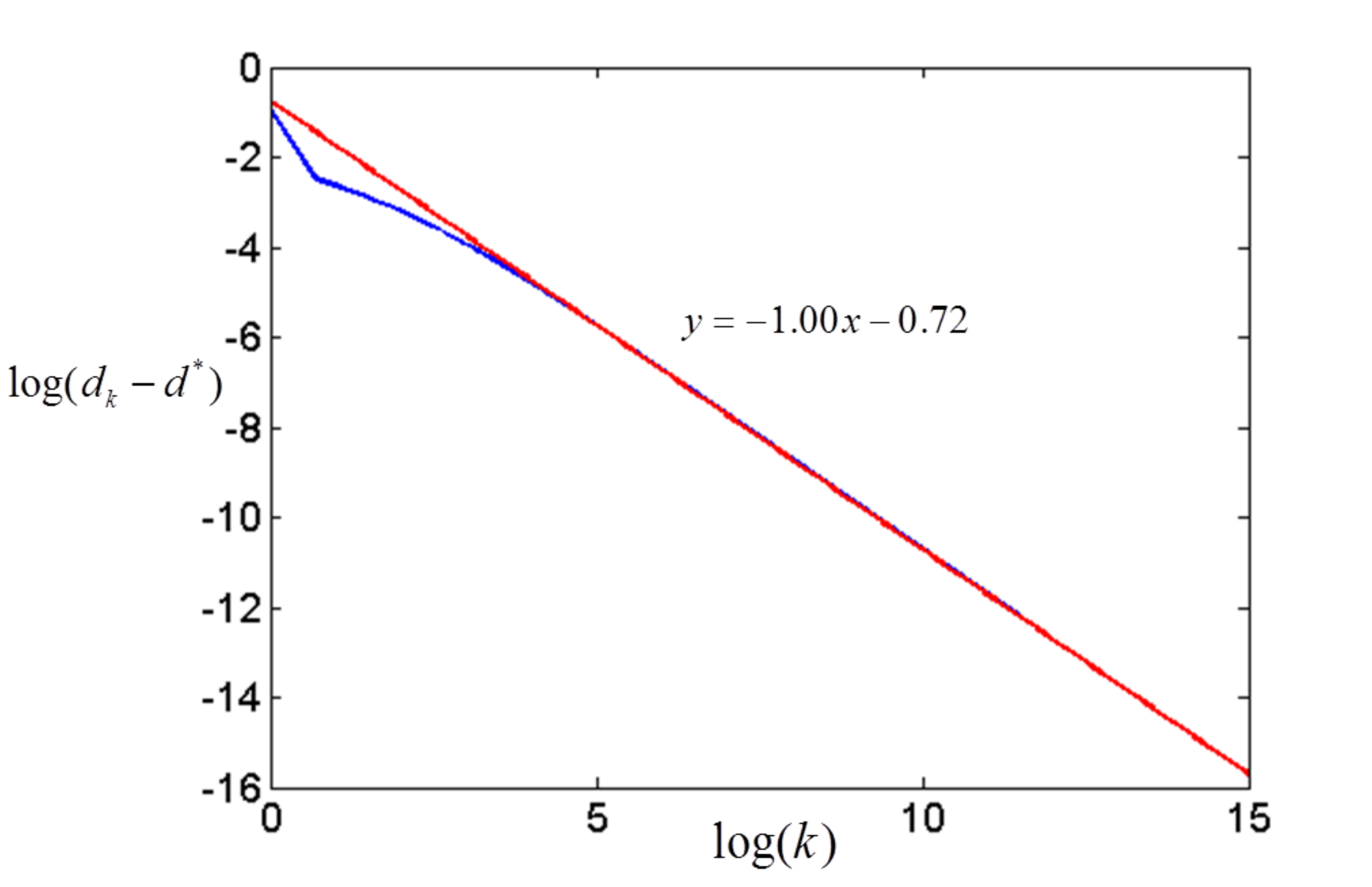} \caption{\textbf{Numerical convergence for the rectangle.} Blue: plot of $\log(d_{k}-d^{*})$ versus $\log(k)$ for the rectangle for $10^{5}$ iterations. Red: best linear fit. Except for the first iterations, both curves are virtually indistinguishable.}
\label{conv_rect}
\end{figure}

Let us first test the accuracy of bound (\ref{bound}). Remember that
such a bound only applies when $\bar{r}\not\in S$. Identifying the
point $\bar{s}^{*}$ with the origin of coordinates, take $\bar{r}=(0,1)$
and set the upper vertices of the rectangle to be $(\pm1,0)$. Our
initial point will be $s_{0}=(0.5,-1)$. Figure \ref{conv_rect} shows
a numerical plot of $\log(d_{k}-d^{*})$ versus the logarithm of the
number of iterations of Gilbert's algorithm. The plot can be well fitted by an
equation of the form $\log(d_{k}-d^{*})=-1.00\log(k)-0.72$. These
results are thus in great agreement with the theoretical upper bound
(\ref{bound}).

\begin{figure}[H]
\centering \includegraphics[width=12cm]{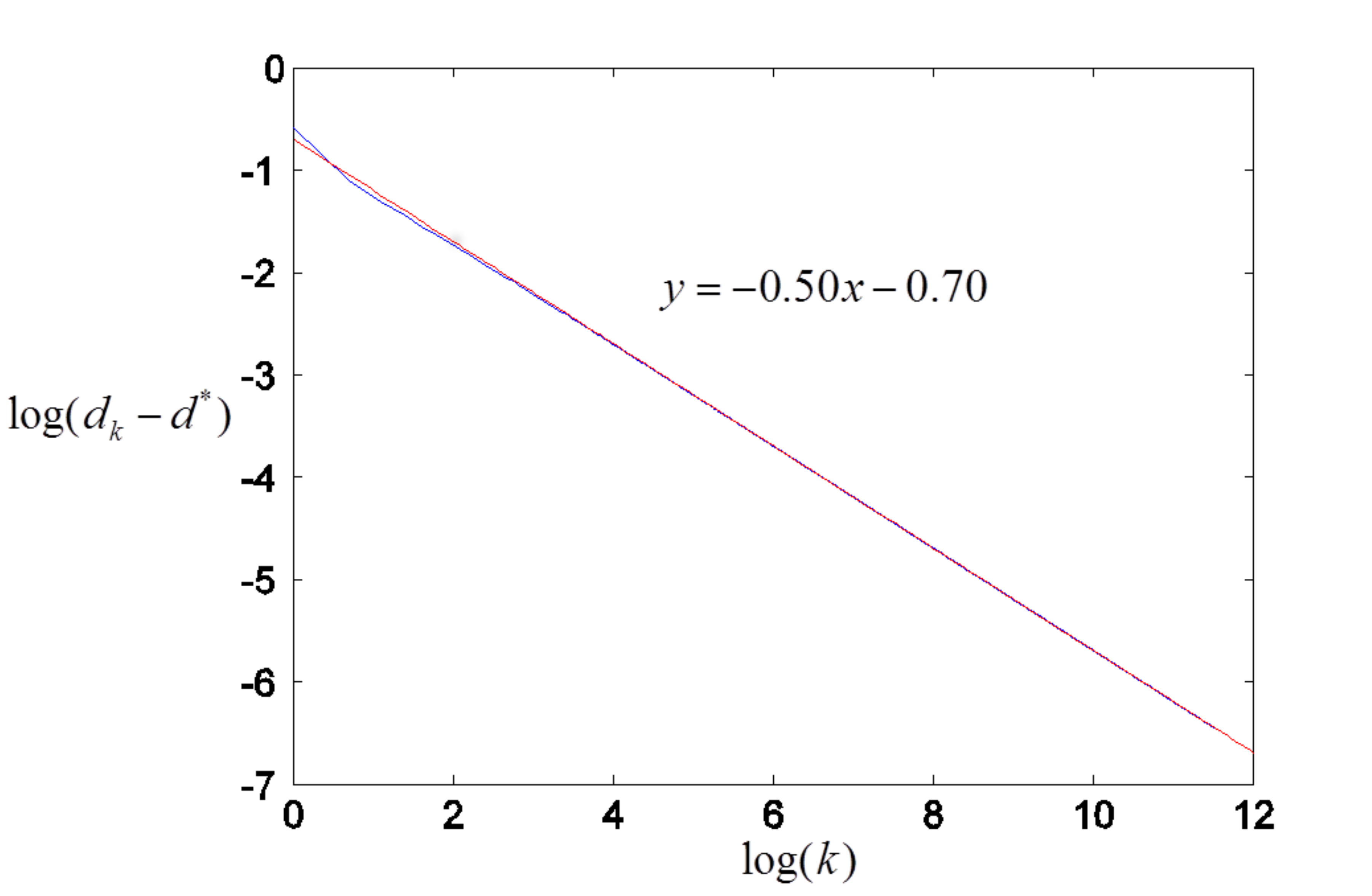} \caption{\textbf{Numerical convergence for the rectangle.} Blue: plot of $\log(d_{k}-d^{*})$ versus $\log(k)$ for the rectangle for $10^{5}$ iterations. Red: best linear fit.}
\label{conv_rect2}
\end{figure}

To test bound (\ref{bound_malo}), we will use the same settings as
before, with the exception of point $\bar{r}$, that we will set at
$\bar{r}=(0,0)$, i.e., the point will be laying on the center of
the side of the rectangle. The results are shown in Figure~\ref{conv_rect2}.
The resulting curve fits very well the formula $\log(d_{k}-d^{*})=-0.50\log(k)-0.70$,
evidencing that the scaling of bound (\ref{bound_malo}) in worst-case
scenarios is correct.

\begin{figure}[H]
	\centering \includegraphics[width=12cm]{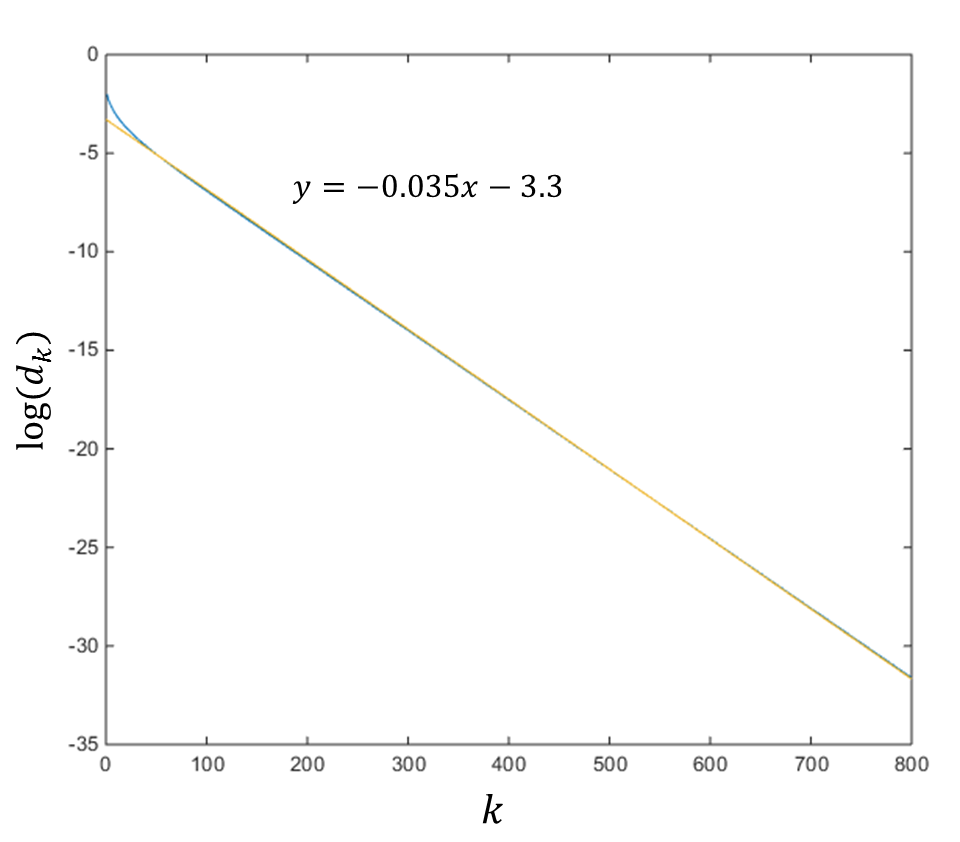} \caption{\textbf{Numerical convergence for inner points of a rectangle.} Blue: plot of $\log(d_{k})$ versus $k$. Yellow: best linear fit.}
	\label{conv_rect3}
\end{figure}

Finally, we test the exponential bound for inner points. Choosing the other two vertices of the rectangle to lie in $(1,-1)$ and $(-1,-1))$, respectively, and taking the inner point $\bar{r}=(-0.5, -0.1)$, we obtain the plot of Figure \ref{conv_rect3}.

\noindent Again, the match between the theory and practice is remarkable.

As we saw in Section \ref{sec: curved_sets}, the above general bounds
are not tight when applied to sets $S$ with a smooth surface: studying
the convergence of the algorithm in less `polyhedric' sets is hence
important. An interesting (curved) set where the convergence rate
can be easily calculated is a circle.

\begin{figure}[H]
\centering \includegraphics[width=12cm]{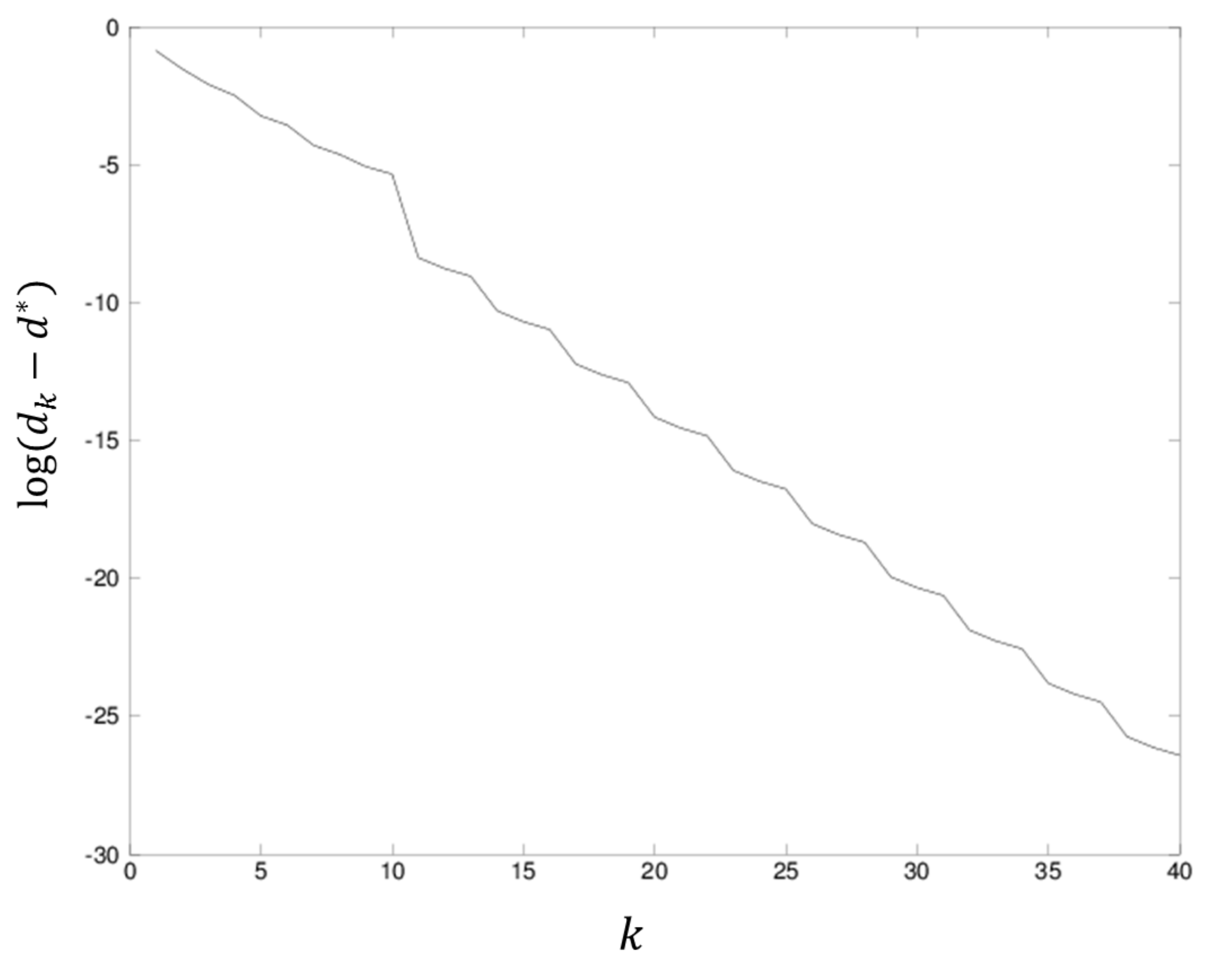} \caption{\textbf{Numerical convergence for the circle.} The plot reveals exponential convergence for curved surfaces.}
\label{circle}
\end{figure}

Setting the origin of coordinates in the center of a circle of radius
1, we placed the external point at $\bar{r}=(0,1.3)$, and took $s_{0}=(1,0)$
to be the initial input for Gilbert's algorithm. The outputs of Gilbert's algorithm are
plotted in Figure \ref{circle}, and show an unequivocal exponential
convergence.

\begin{figure}
\includegraphics[bb=0bp 6cm 595bp 842bp,clip,scale=0.65]{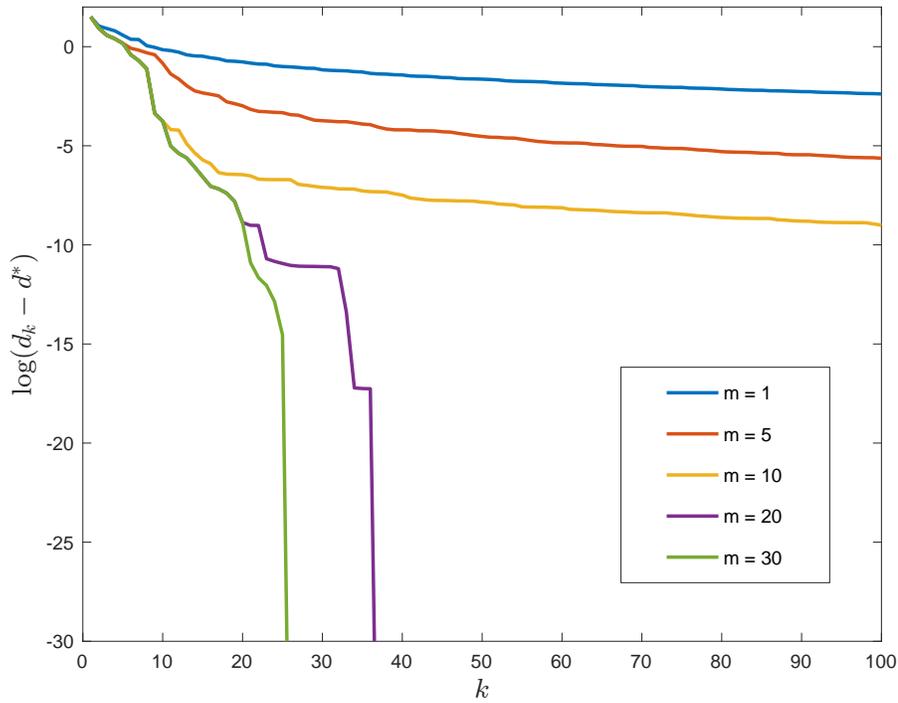}\caption{Improved convergence by storing $m=1,5,10,20$ \& $30$ optimization points. The set is defined as the convex hull of $2^{39}$points in dimension $n=20^{2}.$ The scale is $\log(d-d^{*})$ on the left axis and the number of iterations, $k$, along the bottom and shows exponential convergence for $m=20$ and $m=30$.}
\label{fig: with memory convergence}
\end{figure}

Figure \ref{Fig: rectangle} shows how storing previous optimization
points can improve convergence. After only two iterations, Gilbert's algorithm with a memory buffer of $m=2$ optimization points finds $\bar{s}^{*}$;
contrasting with the polynomial convergence fitted as $\log(d_{k}-d^{*})=-1.00\log(k)-0.72$.
Figure \ref{fig: with memory convergence} shows a further example
of this dramatic improvement in the convergence for a set defined
as the convex hull of $2^{39}$ points in dimension $n=20^{2}$. For
sufficiently large values of approximately $m\ge20,$ the convergence mimics
that of a curved set and after $k=O(n)$ iterations returns $\bar{s}^{*}$
as it is contained in $\textrm{span}\left(L_{opt}\cup\{\bar{s}_{k}\}\right)$.

\section{Applications \label{sec:Applications}}

%In addition to the cost of each oracle call, the time complexity depends on a property of the boundary of $S.$ Gilbert's algorithm can become slow when encountering a flat face of high dimension on the boundary. We overcome this problem by storing previous optimization points which leads to a trade-off between the time cost of each algorithmic step and the number of iterations. We discuss this trade-off further in Section?.

%\subsection{Bell inequalities \label{sub:Bell-inequalities}}

The original motivation for this paper comes from quantum information theory, where an important primitive consists in deciding whether a vector of experimental data measured by two or more experimentalists is classical, in the sense that it admits a local hidden variable model~\cite{bell64}. The set of all classical vectors forms a convex set whose extreme points correspond to local deterministic strategies. The separation problem for this set is, however, NP-hard when we identify the input size of the problem with the number $n$ of measurement settings available to each experimenter \cite{bellNPhard}.

We find that our algorithm allows us to solve instances of WSEP of input size $n=42$. Our algorithm thus improves on the works of Gondzio et al.~\cite{gondzio12,gondzio+12} and Schwartz et al.~\cite{schwarz}, where the cases $n=13$, $n=20$ could not be respectively overcome. Note that the time complexity of the best algorithm known for the strong optimization problem scales as $O(2^{n})$; for $n=42$ it is therefore impractical to call it more than a few times. Hence, following Section \ref{donJose}, we used a heuristic method to carry out all linear optimisations but the last one. Independently of our work, Montina and Wolf  have also used Gilbert's algorithm together with a heuristic to solve large instances of WSEP in the locality problem \cite{montina+16}.

In the next section, we briefly introduce the notions of Bell correlations and EPR steering correlations, where we restrict ourselves to joint terms in the correlations (i.e. neglect local marginals). The first one gives rise to the device-independent framework of quantum correlations (neither parties are trusted)~\cite{bellreview}, whereas the other one defines a semi-device-independent framework (Alice's side is untrusted and Bob's side is trusted)~\cite{steeringreview}.
After the introduction we discuss three applications: In Sec.~\ref{nonlocwerner}, the nonlocality of two-qubit Werner states are investigated, where we improve the known upper bound of the critical visibility and correspondingly beat the known lower bounds on the Grothendieck constant of order 3 (denoted by $K_G(3)$). In Sec.~\ref{nonlocghz}, we study the nonlocality of noisy three-qubit GHZ states, where we provide a Bell inequality which improves the noise tolerance of this state. Finally, in Sec.~\ref{steeringwerner}, we discuss the EPR-steerability of the two-qubit Werner state using a special configuration of sharp qubit observables. Our obtained critical value for steerability also defines a bound on the joint measurability of the noisy version of these observables.

\subsection{Bell and EPR steering correlations}
\label{BellEPR}

We stick to the so-called correlation scenario (two parties with two measurement outcomes considering only joint correlation terms). A behavior (or a correlator) is a point $Q\in\R^n\times\R^n$, where $n$ stands for the number of measurement settings per party. Classical behavior is defined by the convex hull of all deterministic behaviors (vertices), which form the so-called correlation polytope $\mathcal{C}$. For a given $n$, a vertex labeled by $\lambda$ is defined by a specific assignment of $a_x=\pm1$, $x=1,\ldots,n$ and $b_y=\pm1$, $y=1,\ldots,n$, such that the corresponding vertex is given by an $n\times n$ matrix $D_{\lambda}$ with entries $D_{\lambda}(x,y)=a_xb_y$. This amounts to $2^{2n}/2$ distinct vertices (or deterministic strategies). The $(1/2)$ factor comes from the fact that the inversion $a_x\rightarrow -a_x$, $b_x\rightarrow -b_x$ for all $x=1,\ldots,n$ defines the same vertex. Any classical behavior is a convex combination $\sum_{\lambda} q(\lambda)D_{\lambda}$ with positive weights $q(\lambda)$.

On the other hand, in quantum theory a correlation point $Q$ may not be expressed as $\sum_{\lambda} q(\lambda)D_{\lambda}$ according to Bell's theorem~\cite{bell64}. In general, the entries of $Q$ are given by the formula
\begin{equation}
\label{Q2}
Q(x,y)=\tr(\rho A_x\otimes B_y),
\end{equation}
where $\rho$ is an arbitrary state on $\mathbb{C}^d\otimes\mathbb{C}^d$  (i.e. a trace 1 positive matrix in dimension $d^2$) , whereas $A_x\in\mathbb{C}^d$ and $B_y\in\mathbb{C}^d$ are traceless dichotomic observables (i.e. $\tr A_x = \tr B_y = 0$, $A_x^2=B_y^2=\id$) acting on Alice's and Bob's system, respectively.

A two-qubit Werner state is as follows~\cite{werner89}
\begin{equation}
\label{werner}
\rho_W(v) = v \ket{\psi_-}\bra{\psi_-} + (1- v )\frac{\one}{4},
\end{equation}
where $\ket{\psi_-}$ is the two-qubit singlet $(\ket{01}-\ket{10})/\sqrt 2$. The measurements are traceless qubit observables $A_x=\vec a_x\cdot\vec\sigma$ and $B_y=\vec b_y\cdot\vec\sigma$, where
$\vec a_x$ and $\vec b_y$ are unit vectors in $\R^3$ (the so-called Bloch vectors) and $\vec{\sigma}$ denotes the vector of Pauli matrices $\vec\sigma=(\sigma_1,\sigma_2,\sigma_3)$ with components $\sigma_1=\ket{0}\bra{1}+\ket{1}\bra{0}$, $\sigma_2=-i\ket{0}\bra{1}+i\ket{1}\bra{0}$ and $\sigma_3=\ket{0}\bra{0}-\ket{1}\bra{1}$. In the above formula~(\ref{werner}), $0\le v\le 1$ is the visibility of the Werner state and the entry $(x,y)$  of the correlation point $Q(v)$ is simply given by the dot product
\begin{equation}
\label{Qwerner}
Q(x,y)(v)=-v\vec a_x\cdot\vec b_y
\end{equation}
due to Eq.~(\ref{Q2}). The value $v^{(n)}$ of the critical $v$ for which $Q(x,y)$ becomes a classical behavior no matter the choice of the unit vectors $\vec a_x$ and $\vec b_y$ is unknown for $n>4$. In particular, the critical value in case of $n$ going to infinity, $v^*=\lim_{n\rightarrow\infty}v^{(n)}$, is known to be equal to $1/K_G(3)$. However, its exact value is still unknown. Earlier bounds from the literature imply $1.4176\le K_G(3)\le 1.5163$. Note, however, that Krivine's upper bound $1.5163$~\cite{Krivine,acin06} has been recently improved to the value $1.4706$~\cite{flavien}. In this paper, we improve on the lower bound value $1.4176$ \cite{vertesi,hua} establishing a new lower bound value of $1.4261$.

It is straightforward to extend the two-party correlation scenario discussed above to three parties, in which case we pick the noisy Greenberger-Horne-Zeilinger state~\cite{GHZ}
\begin{equation}
\label{ghzstate}
\rho_{GHZ}(p) = p \ket{GHZ}\bra{GHZ} + (1- p)\frac{\one}{8},
\end{equation}
where $\ket{GHZ}=(\ket{001}+\ket{010}+\ket{100})/\sqrt 3$. For a fixed number of $n$ settings per party, the quantum behavior $Q\in\R^n\times\R^n\times\R^n$  is given by its entries $(x,y,z)$
\begin{equation}
\label{QGHZ}
Q(x,y,z)(p)=\tr(\rho_{GHZ}(p) A_x\otimes B_y\otimes C_z)=p\bra{GHZ}A_x\otimes B_y\otimes C_z\ket{GHZ},
\end{equation}
where $A_x$, $B_y$, $C_z$ are traceless dichotomic qubit observables.
Besides, the deterministic behaviors of the three-party correlation polytope for a given $n$ are given by the vertices $D_{\lambda}$ with entries $D_{\lambda}(x,y,z)=a_xb_yc_z$, where $a_x$, $b_y$, $c_z$ may take all possible combinations of $\pm 1$ values, which amounts to $2^{3n}/2$ vertices. Similarly to the case of bipartite Werner states, we ask for a given $n$ the critical value of $p$, denoted by $p^{(n)}$, above which the point $Q(p)$ cannot be expressed as a convex combination of deterministic strategies $\sum_{\lambda} q(\lambda)D(\lambda)$. We are particularly interested in the case of $n\rightarrow\infty$, which we denote by $p^*$. In case of $n=2$, the critical value $p^{(2)}$ is known to be $1/2$ \cite{GHZ}, which gives an upper bound to $p^*$. The best upper bound for $p^*$ so far is attained in Ref.~\cite{palghz} in case of $n=4$ settings per party: $p^{(4)}=0.4960$. In this paper, we decrease this upper bound down to $p^{(16)}=0.4932$ (using 16 measurements per party).

Lastly, we present an application in the EPR steering scenario~\cite{steering} (and see \cite{steeringreview} for a recent review of the field). Here we consider again the case of two-qubit Werner states. However, instead of the fully device-independent Bell scenario, this time we assume a semi-device independent scenario, where only Alice's system is uncharacterized whereas Bob's measurements are assumed to be well defined. In particular, we can assume without loss of generality that Bob performs three Pauli measurements $B_1=\sigma_1$, $B_2=\sigma_2$ and $B_3=\sigma_3$. Plugging these special measurement directions into (\ref{Qwerner}), we get the $Q\in\R^n\times\R^3$ quantum behavior with entries
\begin{align}
\label{Qsteer}
Q_{x1}(v) &= -v\vec a_x\cdot\vec x = -v a_{x}(1)\nonumber\\
Q_{x2}(v) &= -v\vec a_x\cdot\vec y = -v a_{x}(2)\nonumber\\
Q_{x3}(v) &= -v\vec a_x\cdot\vec z = -v a_{x}(3),
\end{align}
where $x=1,\ldots,n$. On the other hand, in the steering scenario (unlike the Bell scenario) the region of classical correlations is defined by a (still convex) set with a continuous number of extremal points. This region is characterized by the convex hull of the following points:
\begin{equation}
\label{Qunsteer}
Q_{xy}=\tr{\left(A_x\sigma_y\ket{\psi}\bra{\psi}\right)},
\end{equation}
where the pure qubit states $\ket{\psi}\in\mathbb{C}^2$ and $A_x=\pm1$ for $x=1,\ldots,n$ and $y=1,2,3$. Note that in this scenario the value $v^{(\infty)}$ is known to be $1/2$ \cite{steeringreview}, although in principle it can only be achieved exactly in the limit of infinitely many measurement settings. Here we find $v^{(30)}\le 0.5058$ via a steering inequality in case of Alice performing measurements in the buckyball configuration. There is a link between EPR steering and jointly measurability~\cite{jm}. In combination with our result, the set of noisy version of the sharp measurements $A_x=v\vec a_x\cdot\vec\sigma$, $x=1,\ldots,30$ in the buckyball configuration are incompatible for $v>0.5058$ and leads to steering when applied to an arbitrary entangled pure state. To the best of our knowledge, the largest set of measurements investigated in the literature corresponds to the dodecahedron, which has 20 vertices and these directions determine $n=10$ settings leading to the visibility $v^{(10)}\le 0.5236$ \cite{jmfinite}. Hence, our bound $v^{(30)}\le 0.5058$ considerably improves on this value as well.

\subsection{Nonlocality of two-qubit Werner states}
\label{nonlocwerner}

Due to Tsirelson~\cite{Tsirelson87}, $v^{(2)}=1/\sqrt 2$. In that case, the correlation polytope is well characterized and the only nontrivial facet corresponds to the CHSH-Bell inequality~\cite{chsh}, whose maximal violation implies the above value $v^{(2)}=1/\sqrt 2$. Up to $n=4$, the correlation polytope $\mathcal{C}$ is completely resolved and two new facets are obtained beyond the CHSH one~\cite{avis}. None of them decreases the critical visibility, hence $v^{(3)}=v^{(4)}=1/\sqrt 2$. For higher $n$ the complexity of the polytope $\mathcal{C}$ grows rapidly, hence a complete characterization becomes challenging even for moderate $n$. For instance, in case of $n=8$ Avis et al.~\cite{avis} gives a lower bound of $37\,346\,094$ to the number of inequivalent facets. Nevertheless, it is known that $v^{(465)}<1/\sqrt 2$ \cite{vertesi}. Using the algorithm described in Sec.~\ref{donJose}, we prove that even $v^{(24)}<1/\sqrt 2$, hence 24 measurement settings are enough to overcome the $1/\sqrt 2$ limit corresponding to the CHSH inequality. The question of the smallest number of measurements $n$ such that $v^{(n)}<1/\sqrt 2$ is still open, however, we conjecture that either $n=24$ is the correct number or it is very close to it. The algorithm in Sec.~\ref{donJose} also allows us to show that $v^{(42)}\le 0.7012$ which in turn shows that $K_G(3)\ge1.4261$ (beating the best bound $K_G(3)\ge1.4176$ currently known). In addition, an example is presented where the measurement vectors $\vec a_x$, $\vec b_y$ feature a nice configuration on the Bloch sphere, associated with the truncated icosahedron (the so-called buckyball configuration), showing that $v^{(30)}\le 0.7030$.

Below we present how the algorithm in Sec.~\ref{donJose} is applied to our problem. Given a fixed measurement configuration with Bloch vectors $\vec a_x$, $\vec b_y$, $x,y=1,\ldots,n$, the method is as follows. We pick a value of $v$ which we wish to beat, say, $v=1/\sqrt 2$. The above Bloch vectors along with $v$ define a quantum point $Q(v)$ with entries $Q(x,y)=-v\vec a_x\cdot\vec b_y$ according to Eq.~(\ref{Qwerner}), which we conjecture to lie (slightly) outside the correlation polytope $\mathcal{C}$. Given $Q(v)$ and the polytope $\mathcal{C}$ we call the algorithm to solve the weak separation problem WSEP, i.e., it outputs a separating hyperplane with norm $W$, such that
\begin{equation}
\label{hyperplane}
w\equiv\max_{L\in\mathcal{C}}{\tr(L W^t)}<\tr(Q(v)W^t),
\end{equation}
where $W^t$ denotes transposition of the matrix $W$. This certifies that $Q(v)$ is indeed outside the polytope and implies the upper bound
\begin{equation}
\label{vn}
v^{(n)}\le \frac{v w}{\tr(Q(v)W^t)}=\frac{w}{\tr(Q(v=1)W^t)}.
\end{equation}

As discussed in Sec.~\ref{donJose}, within each iteration the algorithm uses an approximate Oracle (the description of such a procedure can be found in Appendix A of Ref.~\cite{flavien}), and it is enough to use an exact Oracle in the very last iteration which computes the value of $w$ correctly via a brute force search. Note that in the exact evaluation of the left-hand side of Eq.~(\ref{hyperplane}), it suffices to compute the maximum over all $2^n$ deterministic strategies of Bob (which can be further divided by 2 due to the absolute sign):
\begin{equation}
\label{detBob}
w=\max_{a_x,b_y=\pm1}\sum_{x=1}^n\sum_{y=1}^n{W_{x,y}a_x b_y}=\max_{b_y=\pm1}\sum_{x=1}^n\sum_{y=1}^n{\left|W_{xy}b_y\right|}
\end{equation}
In Matlab, fast matrix manipulation allows us to evaluate parallel the right-hand side expression in several $b_y$ variables. Furthermore, the usage of an Nvidia 4GB graphics card in our standard desktop PC could still boost the speed with roughly a factor of 3. Eventually, this allowed us to compute the $w$ value for $n=42$ settings in (\ref{detBob}) within one week.

The above described method is optimal for a fixed configuration, however, it is expected to arrive at better upper bounds on $v^{(n)}$ by varying the measurement directions $\vec a_x$ and $\vec b_y$ as well. In the original formulation of the algorithm in Sec.~\ref{donJose} this was not possible, however, we can call the algorithm in an iterative manner such that it can search for optimized measurements as well. It works as follows:

\begin{enumerate}
\item Fix $n$. Generate initial Bloch vectors $\vec a_x$ and $\vec b_y$, $x,y=1,\ldots,n$ (either randomly or corresponding to some nice geometric configuration). Also, set $v=1$.
\item Construct $Q$ point with coordinates $Q(x,y)=-v\vec a_x\cdot\vec b_y$.
\item Given polytope $\mathcal{C}$ and point $Q$, run the algorithm in Sec.~\ref{donJose} to get a hyperplane with norm $W$ and a value $w$ of the witness.
\item Maximize the Bell witness $W$ within the two-qubit quantum set, thus obtaining a point $\tilde Q$. This can be done either with a heuristic search or using an SDP~\cite{Boyd+04} implementing the zeroth level of the NPA hierarchy \cite{npa}. In the latter case, however, we also have to ensure that the $\tilde Q$ is realizable with two qubits.
\item Decrease $v$ slightly and construct the point $Q=v\tilde Q$.
\item Go to step 3 until $w<\tr(QW^t)$.
\end{enumerate}

 Both for $n=24$ and $n=42$, we run the iterative procedure above optimizing the measurements as well. Whereas in case of $n=30$, we chose the 60-vertex truncated icosahedron as the fixed measurement configuration both for Alice and Bob (note that the vertices of this geometry possess an inversion symmetry, and it is enough to choose only half of the vertices). In all the above cases, we run the algorithm in step 3 without memory buffer, which makes this step very fast (although, the number of iterations may become bigger). In a supplementary Mathematica file added to the arXiv submission, we give detailed results of the final measurement directions, the Bell matrices $W$ and the corresponding local bound $w$. In case of $n=30$, ten million iterations were used in Gilbert's algorithm of Sec.~\ref{donJose} by setting $v=0.71$. In cases of $n=24$ and $n=42$, we decreased $v$ from $v=1$ down to $v=0.71$ with $\delta v=0.001$ step size in the above described iterative manner, running the algorithm in step 3 ten thousand times. The following upper bounds on the critical visibility of the Werner states were obtained: $v^{(24)}\le 0.7070$, $v^{(30)}\le 0.7030$, and $v^{(42)}\le 0.7012$. To the best of our knowledge, this last value provides the best lower bound $1/0.7012\simeq1.4261$ to $K_G(3)$.

\subsection{Nonlocality of noisy GHZ states}
\label{nonlocghz}

We use Gilbert's algorithm to improve on the smallest upper bound of 0.4960 on $p^*$ in case of the noisy three-qubit GHZ state~(\ref{ghzstate}). To this end, let us arrange the qubit observables $A_x$, $B_y$, $C_z$ in the plane as follows
\begin{align}
A_x &= \cos\theta^a_x\sigma_1 + \sin\theta^a_x\sigma_2\nonumber\\
B_y &= \cos\theta^b_y\sigma_1 + \sin\theta^b_y\sigma_2\nonumber\\
C_z &= \cos\theta^c_z\sigma_1 + \sin\theta^c_z\sigma_2.
\end{align}
For a fixed number of $n$ settings per party, the quantum point $Q(p)$  is given by its entries
\begin{equation}
\label{QGHZ2}
Q(x,y,z)(p)=p\bra{GHZ}A_x\otimes B_y\otimes C_z\ket{GHZ}=p\cos\left(\theta^a_x+\theta^b_y+\theta^c_z\right),
\end{equation}
Let us in particular choose $n=16$ and $\theta_i^a=\theta_i^b=\theta_i^c=\pi(i-1)/n$ for $i=1,\ldots,n$, resulting in
\begin{equation}
\label{QGHZ3}
Q(x,y,z)(p)=p\cos\left(\frac{(x+y+z-3)\pi}{n}\right),
\end{equation}
where $x,y,z=(1,\ldots,n)$.

Similarly to Sec.~\ref{nonlocwerner}, we run Gilbert's algorithm in Sec.~\ref{donJose} with the above $Q(p)$ setting $p=0.4960$ (i.e., the value we wish to beat) using 10 million iterations. The results along with the Bell witness $W$ and the local bound $w$ are summarized in a supplementary Mathematica file. Plugging these values into the respective formula~(\ref{vn}):
\begin{equation}
p^{(n)}\le \frac{w}{\sum_{x,y,z}Q(x,y,z)(p=1)W(x,y,z)},\nonumber
\end{equation}
we get the upper bound $p^{(16)}\le0.4932$, which defines an upper bound to the critical value $p^*=\lim_{n\rightarrow\infty}p^{(n)}$ too.

Let us mention, however, that the use of modified Gilbert's algorithm is not restricted to Bell scenarios with correlation polytopes. It can be used in any Bell scenario (more than two parties with more than two outputs including marginal terms as well). Moreover, we expect Gilbert's algorithm to be useful in other tasks as well, such as quantum communication complexity~\cite{QCC} or dimension witnesses~\cite{dimwit} (given an efficient characterization of dimension restricted quantum correlations~\cite{finitedim}). Another potential application is the case of EPR steering for which we show benchmark results in the next subsection.

\subsection{EPR steering of two-qubit Werner states}
\label{steeringwerner}

We use the modified Gilbert's algorithm with memory buffer in Sec.~\ref{donJose} and find an upper bound of $v^{(30)}\le0.5058$ on the steerability limit of the two-qubit Werner state for Alice's measurements carried out in the buckyball configuration.
In particular, we fix $v=0.51$ and generate the $30\times 3$ quantum point $Q(v)$ in Eq.~(\ref{Qsteer}) using the buckyball configuration for Alice's unit vectors $\vec a_x$, $x=1,\ldots,30$. Given this point $Q(v)$ and the description of the convex unsteerable set~(\ref{Qunsteer}), we run the modified Gilbert's algorithm with memory buffer to find the witness $W$ which separates $Q(v)$ from this set. In case of $n=30$, the Oracle to be solved
\begin{equation}
\label{US}
w=\max_{a_x=\pm 1} \lambda_{max}\left(\sum_{x=1}^n\sum_{y=1}^3W_{x,y}a_x\sigma_y\right),
\end{equation}
where $\lambda_{max}(M)$ denotes the largest eigenvalue of a matrix $M$, is too costly, hence similarly to Secs.~\ref{nonlocwerner} and \ref{nonlocghz} we use an approximate Oracle to compute the maximum in Eq.~(\ref{US}). This is done via a see-saw type iterative search as follows:
\begin{enumerate}
\item Choose random assignments $\{a_x=\pm 1\}_x$, $x=1,\ldots,n$ for Alice's deterministic strategy.
\item Construct the matrix $M=\sum_{x=1}^n\sum_{y=1}^3W_{x,y}a_x\sigma_y$.
\item Compute the eigenvector $\ket{\psi}$ corresponding to the larger eigenvalue $\lambda_{max}$ of $M$.
\item Set $a_x=+1$  if $\bra{\psi}\left(\sum_{y=1}^3W_{x1}\sigma_y\right)\ket{\psi}>0$, otherwise set $a_x=-1$ for all $x=1,\ldots,n$.
\item Go back to step 2 until convergence of $\lambda_{max}(M)$ is reached.
\end{enumerate}
In the particular implementation, we run the above method 100 times and pick the largest final $\lambda_{max}(M)$, which gives an even more reliable approximation of $w$ in formula~(\ref{US}).

Table~\ref{steeringmemory} shows the performance of the Gilbert algorithm of Sec.~\ref{donJose} using different memory sizes $m$ and number of iterations $k$ in terms of distances $d_{k,m}$ of $Q(v)$ from the unsteerable set of points. In complete agreement with the findings of Sec.~\ref{sec: Numerics}, the use of memory speeds up convergence considerably. For instance, in case of $m=100$, convergence of $d_{k,m}$ is already attained up to 7th digit for $k=1000$ iterations. However, without memory (i.e. the case of $m=1$), even $k=100\,000$ is far from reaching convergence. The $W$ steering witness matrix (after rescaling and truncation) corresponding to ($m=100$, $k=100\,000$) is given in a supplementary Mathematica file, which also provides the unsteerable value $w=\sqrt{4166724363}$ defined by Eq.~(\ref{US}) and the maximal quantum value
\begin{equation}
\tr(Q(v=1)W^t)=\frac{632541 + 282885\sqrt{5}}{\sqrt{58 + 18\sqrt{5}}}\nonumber
\end{equation}
of the witness. According to Eq.~(\ref{vn}), we then get the upper bound
\begin{equation}
v^{(30)}\le \frac{w}{\tr(Q(v=1)W^t)}\simeq 0.5058\nonumber
\end{equation}
on the steerability limit of the two-qubit Werner state for the buckyball measurement configuration.

\begin{table}[t]
\label{steeringmemory}
\centering
\begin{tabular}{|l|lll|}
\hline
$d_{k,m}$& $m=1$ & $m=10$ & $m=100$\\
\hline
%$k=100$   &   $0.259936339206$  &   $0.13517134576039$ & $0.07268149358$\\
$k=100$   &   $0.259\,936\,339$  &   $0.135\,171\,345$ & $0.072\,681\,493$\\
\hline
%$k=1000$   &   $0.09214329493012$  &   $0.04547605552560$ & $0.02330135821$\\
$k=1000$   &   $0.092\,143\,294$  &   $0.045\,476\,055$ & $0.023\,301\,358$\\
\hline
%$k=10\,000$   &   $0.040925438758136$  &   $0.02566970202809$ & $0.0233013096068$\\
$k=10\,000$   &   $0.040\,925\,438$  &   $0.025\,669\,702$ & $0.023\,301\,309$\\
\hline
%$k=100\,000$   &   $0.026695317591082$  &   $0.02358076285528$ & $0.023301309450$\\
$k=100\,000$   &   $0.026\,695\,317$  &   $0.023\,580\,762$ & $0.023\,301\,309$\\
\hline
\end{tabular}
\caption{Distance $d_{k,m}$ of $Q$ from the set of unsteerable points, where $k$ is the number of iterations and $m$ is the size of the memory buffer.}
\end{table}

\section*{Appendix A: computation of $\epsilon_k$}
The purpose of this Appendix is to prove that the value of $\epsilon_k\in [0,1]$ that makes the norm of the right-hand side of eq. (\ref{defin}) minimal is given by eq. (\ref{formulita_epsi}).

First, note that $\bar{d}_{k+1}$ can be decomposed as

\be
\bar{d}_{k+1}=\bar{v}_k^\perp+\left(\frac{(\bar{d}_k\cdot\bar{v}_k)}{\bar{v}_k^2}+\epsilon_k\right) \bar{v}_k,
\label{defin2}
\ee
\noindent where $\bar{v}_k^\perp\cdot\bar{v}_k=0$. By varying the value of $\epsilon_k$, we must minimize the coefficient $(\frac{(\bar{d}_k\cdot\bar{v}_k)}{\bar{v}_k^2}-\epsilon_k)^2$. Now, $\bar{d}_k\cdot \bar{v}_k\leq 0$, with equality iff $\bar{d}_k=\bar{d}^*$. Indeed, suppose that $\bar{d}_k\cdot (\bar{r}-\bar{s})\geq \bar{d}_k\cdot\bar{d}_{k+1}'=\bar{d}_k^2$, then we have that

\be
(\bar{d}^*-\bar{d}_k)^2=(\bar{d}^*)^2+d_k^2-2\bar{d}^*\cdot \bar{d}_k\leq (d^*)^2-d_k^2,
\ee

\noindent which only makes sense if $\bar{d}_k=\bar{d}^*$.

It follows that, unless $\bar{d}_k=\bar{d}^*$, the value of $\epsilon_k$ minimizing the norm of the right hand side of eq. (\ref{defin2}) will be greater than $0$. Clearly, if $\frac{(\bar{d}_k\cdot\bar{v}_k)}{\bar{v}_k^2}\geq -1$, then the optimal value is $\epsilon_k=-\frac{(\bar{d}_k\cdot\bar{v}_k)}{\bar{v}_k^2}$, in which case $\vec{d}_{k+1}=\bar{v}_k^\perp$. Otherwise, for all values of $\epsilon\in [0,1]$, the coefficient $(\frac{(\bar{d}_k\cdot\bar{v}_k)}{\bar{v}_k^2}+\epsilon)$ will be non-positive, and hence the value of $\epsilon$ minimizing the aforementioned coefficient -and so the norm of the right hand side of eq. (\ref{defin2})- is $\epsilon_k=1$.

\section*{Appendix B: Stopping criterion for the Separation problem \label{sec:App: Stopping}}

Every call to OPT in step 2 of Gilbert's algorithm returns a vector $\bar{d}_{k}'$
such that

\begin{equation}
\bar{d_{k}}\cdot(\bar{r}-\bar{s})\geq\bar{d_{k}}\cdot\bar{d_{k}'},
\end{equation}

\noindent for all $\bar{s}\in S$. Consequently, any iteration $k$
such that $\bar{d_{k}}\cdot\bar{d_{k}'}>0$ can be interpreted as
a proof that $\bar{r}\not\in S$. Indeed, if $\bar{r}\in S$, then
$\bar{d_{k}}\cdot(\bar{r}-\bar{r})=0$, hence contradicting the above
equation.

When should we expect this to happen? Let $d_{k}\leq d^{*}+\delta$.
By Eq. (\ref{eq: witness}), we have that $(\bar{d}_{k}-\bar{d}^{*})^{2}=d_{k}^{2}+(d^{*})^{2}-2\bar{d}^{*}\cdot\bar{d}_{k}\leq d_{k}^{2}-(d^{*})^{2}\leq(d^{*}+\delta)^{2}-(d^{*})^{2}$.

Now,

\begin{equation}
\bar{d}_{k}\cdot\bar{d}_{k}'=\bar{d}^{*}\cdot\bar{d_{k}'}+(\bar{d}_{k}-\bar{d}^{*})\cdot\bar{d_{k}'}\geq(d^{*})^{2}-\sqrt{(d^{*}+\delta)^{2}-(d^{*})^{2}}(D+d^{*}).
\end{equation}

\noindent A sufficient condition for the stopping criterion to hold
is thus that the right hand side of the above equation is positive,
which can be shown equivalent to:

\begin{equation}
\delta<d^{*}\left(\sqrt{\frac{(d^{*})^{2}}{(d^{*}+D)^{2}}+1}-1\right).
\end{equation}

\end{document}